# THE MODERN THEORY OF EVOLUTION FROM THE VIEWPOINT OF STATISTICAL PHYSICS

ALEXEY V. MELKIKH

Ural State Technical University

19 Mira St., 620002 Yekaterinburg, Russia

mav@dpt.ustu.ru

The problem of the rate and mechanisms of biological evolution was considered. It was shown that species could not be formed due to undirected mutations in characteristic times of about one million years. A mechanism of deterministic molecular evolution assuming a directed change of the genome was proposed.

### Introduction

Views on biological evolution have changed considerably in recent decades since much success has been achieved in the study of the structure and functions of the genome (adaptive mutations, mobile genetic elements, epigenetics, the horizontal transport of genes, etc.). Many scientists speak about the change of the evolution paradigm and new genetics (see, for example, [1-3]), while other hold, as before, to neo-Darwinian views [4-12]. Does all this new knowledge about genes lead to a new theory of evolution or represent just an improvement of the Darwinian theory of evolution?

Modern theories do not provide evidence that some other, rather than Darwinian (undirected changes of the genome, selection and the genetic drift), mechanism of evolution operates. Models of the evolution process are needed so as to make estimates and show what mechanism is in play exactly.

Such estimates are absent in both the neo-Darwinian theory [10] and models of the horizontal transport of genes [13-15], epigenetic processes [16-17], adaptive mutations [18-20], etc. Although there is a great number of studies dealing with simulation of evolution of species, neither author calculated the probability that new species can appear by way of undirected changes of the genome. For example, numerical calculations were made in Eigen's model of quasi-species for short nucleotide chains only. The applicability of the model to real genomes (which are about $10^9$ long) has been postulated as something obvious. However, this is just the main problem: the number of possible combinations of nucleotides rises with the growth of their number in the genome, while the number of organisms, on the contrary, generally decreases (on transition from protozoa to higher organisms).

From the physical viewpoint, Darwinian evolution represents an analogue of the Brownian motion in the space of attributes of organisms. In this sense, it is important to estimate characteristic times of evolution. Can new species appear by diffusion in the space of attributes?

## 1. Evaluation of the Probability of New Species Formation in Terms of the Synthetic Theory of Evolution

According to the modern theory of evolution, the main mechanisms by which new species appear are mutations and the horizontal gene transport. It is assumed that mutations in a genome can be not only casual, but frequently represent rearrangements of blocks already available in the genome. At the same time, the basic proposition of the modern theory of evolution is its *nondirectivity*. That is, whatever mutations may be, they are not directed a priori to creation of "good" combinations of nucleotides. The same reasoning applies to processes of the horizontal transport of genes: while being largely non-casual, they are not oriented to creation of "good" genes.

Otherwise, aprioristic information about the structure of the genome of future species would be needed for directed changes in the genome. According to STE, such aprioristic information is unavailable.

To estimate characteristic times necessary for appearance of new species, we shall consider a genome having the following properties (in accordance with main provisions of the modern theory of evolution):

1. A genome represents an ordered set X with elements (nucleotides) $X_1$, …, $X_N$ (N being the total number of nucleotides in the genome). Each element can take one of four values (A, T, G, C). The following operations $P_i$ are applicable to the elements: replacement of individual elements, replacement of blocks of elements, and any rearrangements of blocks and elements. All these processes have different probabilities.

2. The operations cause appearance of a new organism whose survival probability is determined by the genome structure and properties of the environment. An ecological niche (occupied by one species) represents a region of the phase space where a small number of organisms increases in quantity (the reproduction rate is positive). Let us label the set of genomes of organisms in the niche as $L_0$. At the initial moment of time each niche is surrounded by k *empty* nearest niches (the corresponding sets are labeled $L_1…L_k$). Neighboring ecological niches are separated by a region where the reproduction rate is negative.

3. The nearest neighboring species have $N_1$ different nucleotides (the Hemming characteristic distance between nearest neighboring sets is $L_i$). We shall refer to nucleotides, which belong at least to one of the sets $L_1$ … $L_k$, as "good" ones. The characteristic distance between organisms of one species is labeled $N_2$ (the intraspecific distinction) (Fig. 1).

4. Whatever the operations $P_i$ may be, they are not oriented a priori to creation of new species adapted to the environment. Organisms do not know the location of nearest ecological niches, while positions of "good" nucleotides in the genome are unknown a priori.

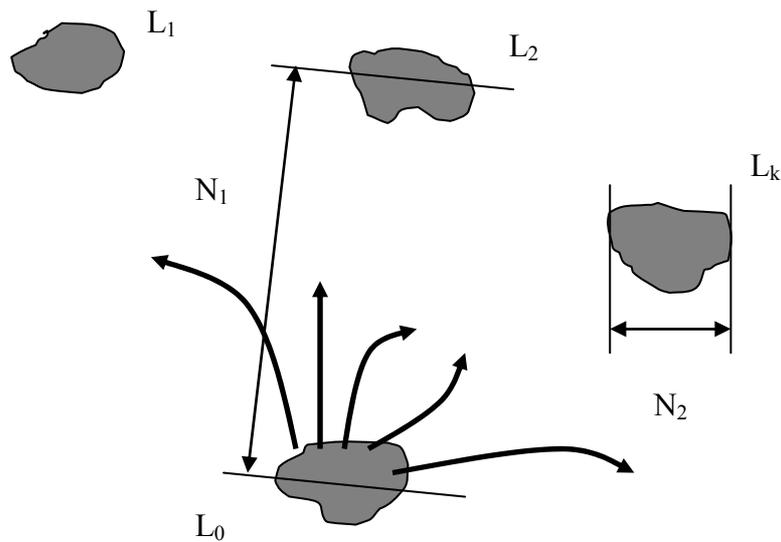

Fig. 1. Ecological niches (colored grey)

These properties coincide in many respects with properties of the quasi-species model [10], but the probability of species formation was not evaluated in terms of the last model for real sizes of genomes. In this sense, the property 4, which forms the basis of STE, has the principal significance.

Let the probabilities that two arbitrarily chosen nucleotides $X_k$ and $X_l$ are "good" (belong to one of neighboring sets $L_1 \ldots L_k$) be equal to $p_k$ and $p_l$ respectively. Does the probability that they are "good" simultaneously depend on the numbers of these nucleotides (their positions in the genome)? If such dependence exists, it will mean the availability of the aprioristic knowledge which nucleotides are "good" and which are not. This is in conflict with the property 4 and changes of the genome prove to be directed. Then the numbering order of nucleotides is not important for calculation of the probability that they are good. Obviously, the notion of "a block" becomes senseless for calculations of this probability: since boundaries of blocks of a neighboring species are unknown a priori, it is not important whether any nucleotides change under the action of operations in a mutually correlated pattern or not.

Since nearest species differ by $N_1$ nucleotides, it is necessary to determine the probability that they will be exactly $N_1$ (considering the distinction between organisms within one species). Furthermore, these nucleotides should be correctly distributed in the genome. Notice that if random walks occur in the space of attributes and the position of neighboring niches is not known beforehand (the darwinian mechanism of evolution), the position of mutated nucleotides in the genome and their structure are unknown either. If such aprioristic knowledge is unavailable, the probability of $N_1$ nucleotides hitting "right" places is

$$W_1 = \frac{(N-N_1)! N_1!}{N!} \frac{1}{3^{N_1}}. \tag{1}$$

The formula (1) is deduced as follows. The probability that the first mutated nucleotide is "right" (i.e. not only hits the place of mutation, but also is the one needed) equals

$$\frac{N_1}{3N}.$$

The probability for the second mutated nucleotide is

$$\frac{N_1 - 1}{3(N-1)}.$$

And for the last

$$\frac{1}{3(N-N_1)}.$$

The general probability is the product of these probabilities and, hence, we have the formula (1).

If an error, which is possible at this stage (the intraspecific distinction), is taken into account, we have:

$$\frac{(N-N_1)!N_1!}{N!}\frac{1}{3^{N_1-N_2}}. \qquad (2)$$

Let us estimate this value using Stirling's formula for large numbers:

$$\ln W_1 = \ln \frac{(N-N_1)!N_1!}{N!}\frac{1}{3^{N_1-N_2}} = (N-N_1)\ln(N-N_1) + N_1 \ln N_1 - N \ln N - (N_1-N_2)\ln 3$$

For example, at $N = 3 \times 10^9$, $N_1 = 0.01N$ and $N_2 = 0.001N$ (for example, see [21-24]) we have

$$\ln W_1 = -0.066 N \approx -2 \times 10^8.$$

Then for $W_1$ we obtain

$$W_1 \approx \exp(-2 \times 10^8).$$

Thus, the probability that mutated nucleotides prove to be exactly the ones needed for formation of a new species is vanishingly small: *new species of organisms could not appear due to undirected mutations.*

Let us estimate separately the probability that the number of mutated nucleotides will fall within the interval

$$N_1 - \frac{N_2}{2} < N_X < N_1 + \frac{N_2}{2}.$$

Let the number of mutated nucleotides $N_X$ have the Gaussian distribution. Since this number is large, this distribution may be viewed as continuous:

$$f(N_X) = \frac{1}{\sqrt{2\pi} N_{2m}} \exp\left(-\frac{(N_X - N_{1m})^2}{2N_{2m}^2}\right).$$

Here $N_{2m}$ represents the characteristic width of the distribution. The probability that the number of nucleotides falls within the specified interval is

$$W_2 = \int_{N_1-N_2/2}^{N_1+N_2/2} \frac{1}{\sqrt{2\pi} N_{2m}} \exp\left(-\frac{(N_X - N_{1m})^2}{2N_{2m}^2}\right) dN_X.$$

What parameters of this distribution should be for W to be a maximum? It is easy to show that the maximum $W_2$ (equal to unity) will be realized when

$$N_{1m} = N_1$$

and $N_{2m}$ tends to zero. In this case, the distribution of nucleotides represents a δ-function. Of course, this assumption is unrealistic and in the nature any such distribution (even if nucleotides are not independent) has a final dispersion. Therefore, if the real distribution of nucleotides is taken into account, the probability of hitting the specified interval decreases still more. Let us make the upper estimate, i.e. assume that exactly

$$N_1 - \frac{N_2}{2} < N_X < N_1 + \frac{N_2}{2}$$

nucleotides change as a result of mutation.

How many attempts are required so that at least one organism reaches any of the neighboring ecological niches? Since positions of elements belonging to the sets $L_1 \ldots L_k$ are unknown, the distance between the sets remains (on the average) the same (according to the property 4, all the operations are not directed) after the first attempt, i.e. can decrease or increase with an equal probability. Therefore, the total number of attempts can be counted taking the number of organisms of a given species ever living on the Earth. They may be both descendants of one organism (transition in a great number of small steps) and descendants of other organisms (parallel transition).

Therefore, we shall multiply the probability that at least one organism gets into a neighboring niche by the number of organisms of the given species ever living on the Earth. Thus, all possible attempts (whatever their factor of multiplication between niches be, the real number of organisms is limited by natural resources) are taken into account. If some organisms do not survive the action of selection, the number will be still smaller.

Even the very probability of the first hit represents a combinatorially small number. It is obvious therefore that real dimensions of populations of any organisms cannot considerably increase this probability.

Finally, the obtained probability is multiplied by the number of nearest ecological niches adjacent to the given niche. This operation gives (considering the assumptions made) the upper estimate of the probability that at least one organism reaches at least one neighboring ecological niche:

$$W = m\frac{t}{T}k\frac{(N-N_1)!N_1!}{N!}\frac{1}{3^{N_1-N_2}}. \qquad (3)$$

Here T is the lifetime of an organism, t is the characteristic time of formation of a new species, m is the number of a population of organisms, and k is the number of ecological niches adjacent to the given niche. Let us also take into account the known relations $N_1 = 0.01N$ and $N_2 = 0.001N$. The $W_1$ estimate (2) suggests that the number W is negligibly small too ($W \approx 10^{-57000000}$).

The estimates disregarded processes of the further conversion of information in the genome (the genome regulation, alternative splicing, etc.). It can be shown however that the probability of formation of new species is of the same order of magnitude if one compares sequences of amino acids in proteins.

Thus, species could not appear due to undirected mutations. Therefore, a considerable part of mutations and operations of the horizontal transport of genes are directed to creation of organisms a priori adapted to new ecological conditions. In this case, the very mechanism of evolution changes drastically: it turns from casual to deterministic.

2. **Genome as a Neurocomputer**

There is a variety of papers, in which a genome is treated as a network of interconnected genes analogous to a neuron network (see, for example, [25-28]). If an analogy is drawn between a genome (or several genomes) and a neuron network, it should be noted that a fundamental property of a neuron network, without which it cannot work, is the presence of two modes: recording and pattern recognition. The work of a neuroprocessor starts from presentation of a pattern. The pattern presentation (input of a primary set of attributes) is realized as follows. At the initial moment of time signals that activate some elements are transmitted via *external links*. A presented pattern is maintained for some time, during which the links "*learn*" (in other words, conductivity of current-carrying links decreases). After the learning procedure, the processor can recognize objects by relating them to some class of objects it learned.

Thus, a neuron network (or similar structures) will not work without *a priori* standards (an initial set of attributes or reference patterns)! In this case, it has nothing to compare with a received signal and the decision as to the "good" or "bad" signal is impossible. If some special assumptions on properties of a "genome" neuron network are not made, this network cannot process signals from the environment and adequately react to them.

In this case, environmental effects will have the character of a random process (in the aforementioned sense) and cannot accelerate evolution. That is, the model proposed in [2] is a particular case of neo-Darwinism.

If it is assumed that a program of genes control appeared casually, we again encounter a contradiction since the probability of formation of this program proves to be vanishingly small. If, thanks to random processes, a genome forms a sequence encoding proteins, which are responsible for shuffling of genes or their parts, there are no reasons to assert that new genes will be "good". The probability of this process should be estimated using the formula (3) for random processes, which gives vanishingly small probabilities of such events.

### 3. Main Ideas of the Deterministic Theory

Of course, the construction of the deterministic theory requires a special discussion, but even now it is possible to formulate basic ideas of this theory (see also [29, 30]).

1. Random processes, which cause changes of the genome, may take place concurrently with deterministic processes representing the purpose-oriented work of molecular machines. Such a controlled change of the genome essentially approaches the morphogenesis. Consequently, new species of organisms will not appear casually, but will result from a deterministic process.

2. Information cannot appear from nothing (such a process would contradict the second law of thermodynamics). Therefore, aprioristic information about new species should be encoded in some structures. Conformational degrees of freedom of proteins presumably may serve as the storage of this aprioristic information. These degrees of freedom represent an additional information resource since only the sequence of protein amino acids, but not the spatial structure of the protein is encoded in genes.

3. Laws of functioning of molecular machines are general for different cell processes. From the viewpoint of statistical thermodynamics of irreversible processes, the work of these machines consists in an efficient conversion of one form of energy to its another form – the cross effect (see also [31]). Operations on genes, which can be a result of the work of molecular machines, may be reduced to several elementary operations, such as identification of a DNA fragment (a protein), cutting of a macromolecule, cross-linking, etc.

## Conclusion

Thus, this study showed that undirected evolution of organisms takes a too long time for appearance of new species (including the case where genes form a complex network). A mechanism of deterministic evolution was proposed. The essence of this mechanism is that possible species of organisms are predetermined by properties of proteins and nucleotides. The structure and chemical properties of nucleotides, amino acids and other substances essential for life are such that changes in a genome, which lead to appearance of new species, become controllable. The formation of new species represents a deterministic process approaching the morphogenesis.